\documentclass[notitlepage,twocolumn,prl,superscriptaddress]{revtex4-1}
\usepackage{amsmath}
\usepackage{changes}
\usepackage{graphicx,placeins}
\usepackage[space]{grffile} %Import files with spaces in names
\usepackage{gensymb}
\usepackage{bm}% bold math
\usepackage{amssymb,amsfonts,amsmath,amsbsy,bm,t1enc,latexsym}
\usepackage{soul}
\usepackage{amssymb} %maths
\usepackage{amsmath} %maths
\usepackage{amsfonts}
\usepackage{amsmath}
\usepackage{amssymb}
\usepackage{graphicx}
\RequirePackage{lineno}
\usepackage{siunitx}
\usepackage{hyperref}       % hyperlinks
\hypersetup{
    colorlinks=true,
    linkcolor=blue,
    filecolor=magenta,      
    urlcolor=cyan,
}

\usepackage[utf8]{inputenc} %useful to type directly diacritic characters

\newcommand{\sym}{\mathrm{s}}
\newcommand{\opa}{\hat{a}}

\newcommand{\ant}{\mathrm{a}}

\newcommand{\OP}[1]{\hat{#1}} %operators
\newcommand{\OPC}[1]{\hat{#1}^\dagger} % adjoint
\newcommand{\OPH}{\OP{H}}
\newcommand{\OPa}{\OP{a}}
\newcommand{\OPCa}{\OPC{a}}
\newcommand{\OPb}{\OP{b}}
\newcommand{\OPCb}{\OPC{b}}
\newcommand{\Jc}{J}
\newcommand{\gc}{g}
\newcommand{\asym}{\mathrm{a}}
\newcommand{\OPas}[1][]{\OP{a}_{\sym#1}}
\newcommand{\OPaas}[1][]{\OP{a}_{\asym#1}}
\newcommand{\OPCas}[1][]{\OPC{a}_{\sym#1}}
\newcommand{\OPCaas}[1][]{\OPC{a}_{\asym#1}}
\newcommand{\opfwm}{\OP{k}}

\begin{document}

\newcommand{\abs}[1]{\left|#1\right|}
\newcommand{\abssq}[1]{\abs{#1}^2}%norm squared
\newcommand{\diff}[1]{\mathrm{d}#1}
\newcommand{\drt}{\tfrac{\mathrm{d}}{\diff{t}}}
\renewcommand{\Re}{\text{Re}}%real part
\renewcommand{\Im}{\text{Im}}%real part

\newcommand{\AS}{A_\mathrm{s}}
\newcommand{\AAS}{A_\mathrm{as}}
\newcommand{\ASmu}{A_{\mathrm{s},\mu}}
\newcommand{\AASmu}{A_{\mathrm{as},\mu}}

\title {Emergent Nonlinear Phenomena in a Driven Dissipative Photonic Dimer}

\author{A.~Tikan$^1$, J.~Riemensberger$^1$, K.~Komagata$^{1,3}$, S.~H\"onl$^2$, M.~Churaev$^1$, C.~Skehan$^1$,  H.~Guo$^{1,4}$, R.~N.~Wang$^1$, J.~Liu$^1$, P.~Seidler$^2$, T.J.~Kippenberg}
\affiliation{Institute of Physics, Swiss Federal Institute of Technology Lausanne (EPFL), CH-1015 Lausanne, Switzerland \\ $^2$IBM Research Europe, Säumerstrasse 4, CH-8803 Rüschlikon, Switzerland \\ $^3$ Present address: Laboratoire Temps-Fréquence, Institut de Physique, Université de Neuchâtel, CH-2000 Neuchâtel, Switzerland \\
$^4$ Present address: Key Laboratory of Specialty Fiber Optics and Optical Access Networks, Shanghai University, 200444 Shanghai, China}

\date{\today}
\email{alexey.tikan@epfl.ch, tobias.kippenberg@epfl.ch}

\maketitle

\begin{figure*}
	\centering
	\includegraphics[width=\linewidth]{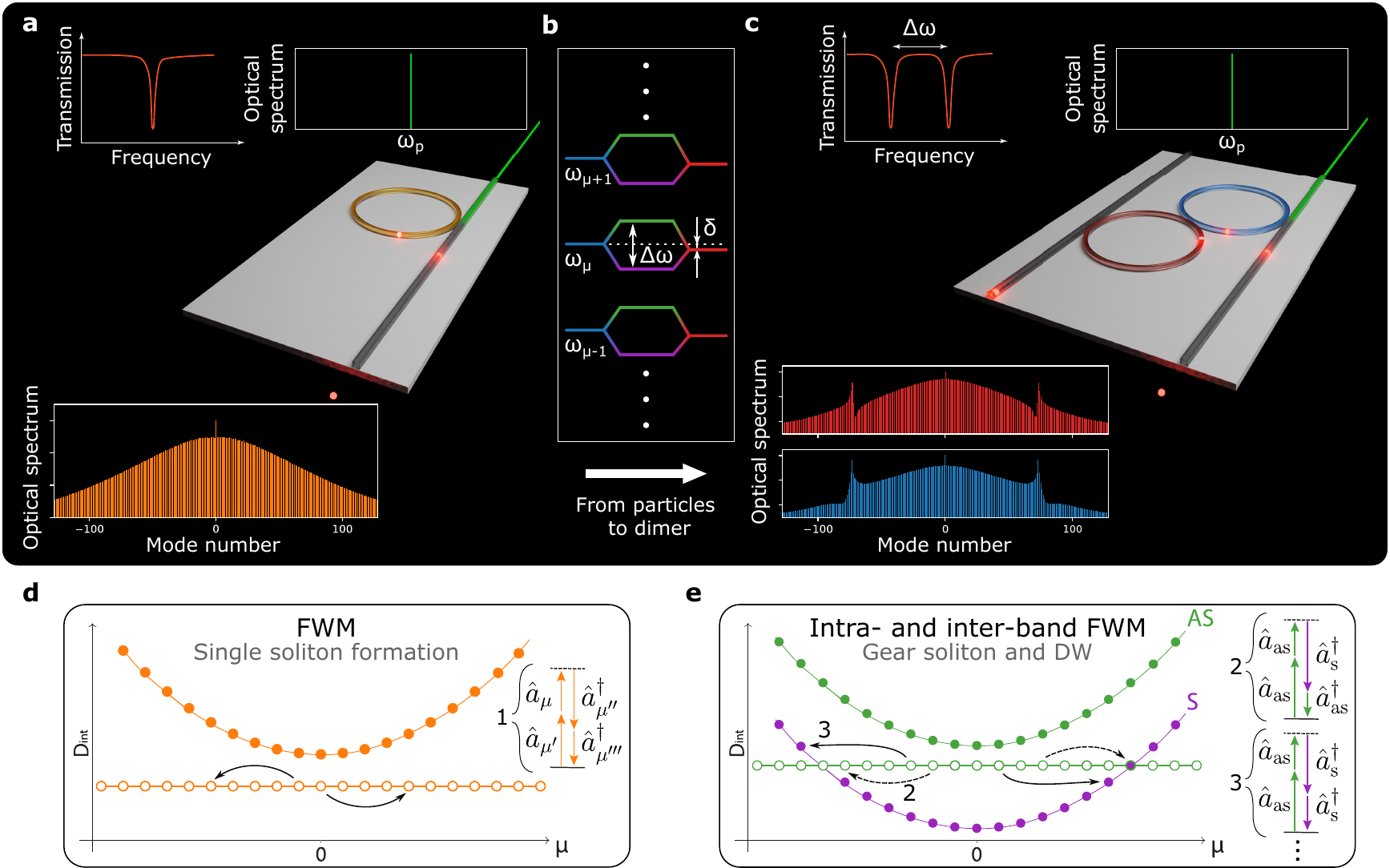}
	\caption{\textbf{Dissipative Kerr Soliton (DKS) formation in a single and coupled ring resonators}. \textbf{(a)} Single resonator (particle) case. Dissipative Kerr soliton generation with a monochromatic pump (green). A resonance in the linear transmission trace is represented by a single Lorentzian dip. The spectrum of a single DKS generated in the resonator is shown by discrete orange lines having a hyperbolic secant profile. \textbf{(b)} Mode splitting upon transition to the dimer case. Resonator modes initially separated by the inter-resonator detuning $\delta$ are hybridized in the dimer configuration and form a two-step ladder with the separation $\Delta\omega= \sqrt{4 J^2 + \delta^2}$. \textbf{(c)} Two strongly coupled resonators (dimer case). Simultaneous spatiotemporal self-organization in both cavities forms the GS (Gear Soliton). The hyperbolic secant spectrum is modified by two symmetrically-spaced Fano shapes. The output spectrum of the pumped resonator is shown with blue lines, while the one of the auxiliary with red ones. Orientation of the Fano shapes depends on the sign of the inter-resonator detuning. \textbf{(d)} Cascaded FWM (Four-Wave Mixing) in the particle case leading to a DKS formation. \textbf{(e}) Novel FWM pathways between the supermodes which take place in the photonic dimer. Odd inter-band (2, dashed line) and even inter-band (3, solid line) FWM pathway leading to the emergence of DWs (Dispersive Waves).}
	\label{fig:1}
\end{figure*}

\textbf{
Emergent phenomena are ubiquitous in nature and refer to spatial, temporal, or spatiotemporal pattern formation in complex nonlinear systems driven out of equilibrium that is not contained in the microscopic descriptions at the single-particle level~\cite{anderson1972more,prigogine1978time,bak1996nature}. Examples range from novel phases of matter in both quantum and classical many-body systems, to galaxy formation or neural dynamics~\cite{schulman1986percolation,kevrekidis2007emergent,zhang2017observation,choi2017observation}. Two characteristic phenomena are length scales that exceed the characteristic interaction length and spontaneous symmetry breaking~\cite{nishimori2010elements,prigogine1968symmetry}. Recent advances in integrated photonics~\cite{liu2018ultralow,Liu2020Photonic} indicate that the study of emergent phenomena is possible in complex coupled nonlinear optical systems~\cite{Jang2018Synchronization,Morichetti2012Coupled,Vasco2019Slow}.
Here we demonstrate that out-of-equilibrium driving of a strongly coupled (‘dimer’) pair of photonic integrated Kerr microresonators~\cite{Zhang2019Electronically}, which at the ’single-particle’ (i.e. individual resonator) level generate well understood dissipative Kerr solitons~\cite{Herr2014Temporal,Kippenberg2018Dissipative}, exhibits emergent nonlinear phenomena. By exploring the dimer phase diagram, we find unexpected and therefore unpredicted regimes of soliton hopping, spontaneous symmetry breaking, and periodically emerging (in)commensurate dispersive waves. These phenomena are not included in the single-particle description and are related to the parametric frequency conversion between hybridized supermodes. Moreover, by controlling supermode hybridization electrically~\cite{miller2015tunable,xue2015normal}, we achieve wide tunability of spectral interference patterns between dimer solitons and dispersive waves. Our findings provide the first critical step towards the study of emergent nonlinear phenomena in soliton networks and multimode lattices.
}

Increasing the number of the components in a dynamical nonlinear system often leads to the appearance of so-called \textit{emergent phenomena}~\cite{bak1996nature} that are accompanied by the violation of underlying symmetries and even microscopic laws~\cite{anderson1972more,prigogine1978time}. Emergent phenomena are omnipresent and most often observed as spontaneous self-organization of spatiotemporal patterns. The formation of galaxies~\cite{schulman1986percolation}, complex neural interaction in the human brain~\cite{chialvo2010emergent} or collective dynamics in Bose-Einstein condensates~\cite{kevrekidis2007emergent} can be understood as arising from emergent phenomena. Recently, the advances in the manipulation of driven dissipative quantum systems have allowed the study of non-equilibrium phases in strongly interacting quantum matter and led to the discovery of time crystals~\cite{zhang2017observation,choi2017observation}. Particularly important characteristic phenomena associated with emergent dynamics are phase transitions~\cite{nishimori2010elements} and symmetry breaking~\cite{prigogine1968symmetry} in complex systems. Their properties are actively studied in different branches of physics including photonics.
Emergent phenomena also occur in driven dissipative nonlinear optical systems out of equilibrium.
For example, the complex self-organization of light in the form of dissipative solitons~\cite{akhmediev2008dissipative} in active nonlinear optical cavities related to the phenomenon of mode-locking~\cite{haus2000mode,grelu2012dissipative}, paved the way for efficient ultrashort pulse generation, that is the foundation of modern frequency metrology~\cite{cundiff2003colloquium}. %Among recent examples, we can highlight the study of spatiotemporal mode-locking, i.e. the mode-locking in the domain of multiple transverse mode families, in addition to the longitudinal one~\cite{wright2020mechanisms,Wright2017Spatiotemporal}.
More recently, spatiotemporal mode-locking, i.e. the mode-locking in the domain of multiple transverse mode families in addition to the longitudinal one~\cite{wright2020mechanisms,Wright2017Spatiotemporal}, has been reported which is an important example of the presence self-organization of light in more complex nonlinear optical systems. 

The discovery of coherent localized light states, Dissipative Kerr Solitons (DKS)~\cite{Kippenberg2018Dissipative}, in continuous wave-driven passive nonlinear microresonators~\cite{Herr2014Temporal,Yi2015Soliton,brasch2016photonic} has heralded a new generation of optical frequency combs that can now be integrated on-chip providing  a universal platform for various applications ~\cite{Liu2020Photonic}.
What makes driven Kerr cavities particularly attractive in the study of nonlinear phases of driven dissipative nonlinear systems, is the ability to map out experimentally the two-dimensional stability chart spanned by laser pump power and laser detuning and, therefore, to explore experimentally the rich and complex phase diagrams of DKS.
Indeed, DKS dynamics has been extensively studied over the last years providing an accurate understanding of new physical phenomena ranging from breathers to soliton crystals~\cite{jang2015temporal,guo2017universal,Lucas2017Breathing,cole2017soliton,cole2018kerr,wang2017universal,karpov2019dynamics}. Also, perturbations to the exact model such as Raman scattering or influence of avoided mode crossings have been largely explored~\cite{karpov2016raman,Herr2014Mode,lobanov2015frequency,xue2015mode,yi2017single,yu2020spontaneous}.  However, all these novel and recently reported underlying dynamical mechanisms can be well explained within the one-dimensional 'single-particle' (i.e. individual resonator) Lugiato-Lefever equation~\cite{Lugiato1987Spatial,Chembo2013Spatiotemporal,Lugiato2018From} and its modifications. Thus, the study of emergent phenomena in driven nonlinear optical systems out of equilibrium has been mostly confined to the cases with comparatively few degrees of freedom.

Here we report the study of \emph{emergent nonlinear phenomena} of the DKSs dynamics beyond the Lugiato-Lefever model by adding another spatial dimension. Our study differs fundamentally from all prior results, which were obtained by including perturbations into the 'single-particle' Lugiato-Lefever. Instead, here we consider dynamics resulting from two \emph{perfect} (i.e. ideal and unperturbed) coupled systems. Investigating DKS formation in the photonic dimer~\cite{Zhang2019Electronically} - a fundamental element of soliton lattices - we show surprisingly rich nonlinear dynamics related to the efficient photon transfer between the dimer supermodes. We study the case when DKSs are generated in both resonators and due to the underlying field symmetry, we refer to these structures as \textit{Gear Solitons} (GSs). We show that the dimer phase diagram features novel states fundamentally inaccessible in single resonators.
We numerically demonstrate the emergence of commensurate dispersive waves (DW) whose periodic enhancement leads to the discretization of the soliton existence range, soliton hopping, as well as an effect of symmetry breaking related to the discreteness of the system. For the first time, we experimentally observe GSs formation in a strongly coupled photonic dimer and probe their non-trivial dynamics. Strikingly and counter-intuitively, we observe that the imperfect mode hybridization (i.e. finite inter-resonator detuning) is required to initiate coherent dissipative structures. Moreover, on a practical level, the combination of a dimer with microresonator tuning, enables electronically control of hybrid dispersive waves. Our results highlight the richness and complexity to be explored in the emergent nonlinear dynamics of multimode resonator lattices.% and at the same time offers a practical route to achieve electrical control over underlying nonlinear processes.

\section{Physical model of coupled Kerr resonators}

\begin{figure*}
	\centering
	\includegraphics[width=0.95\linewidth]{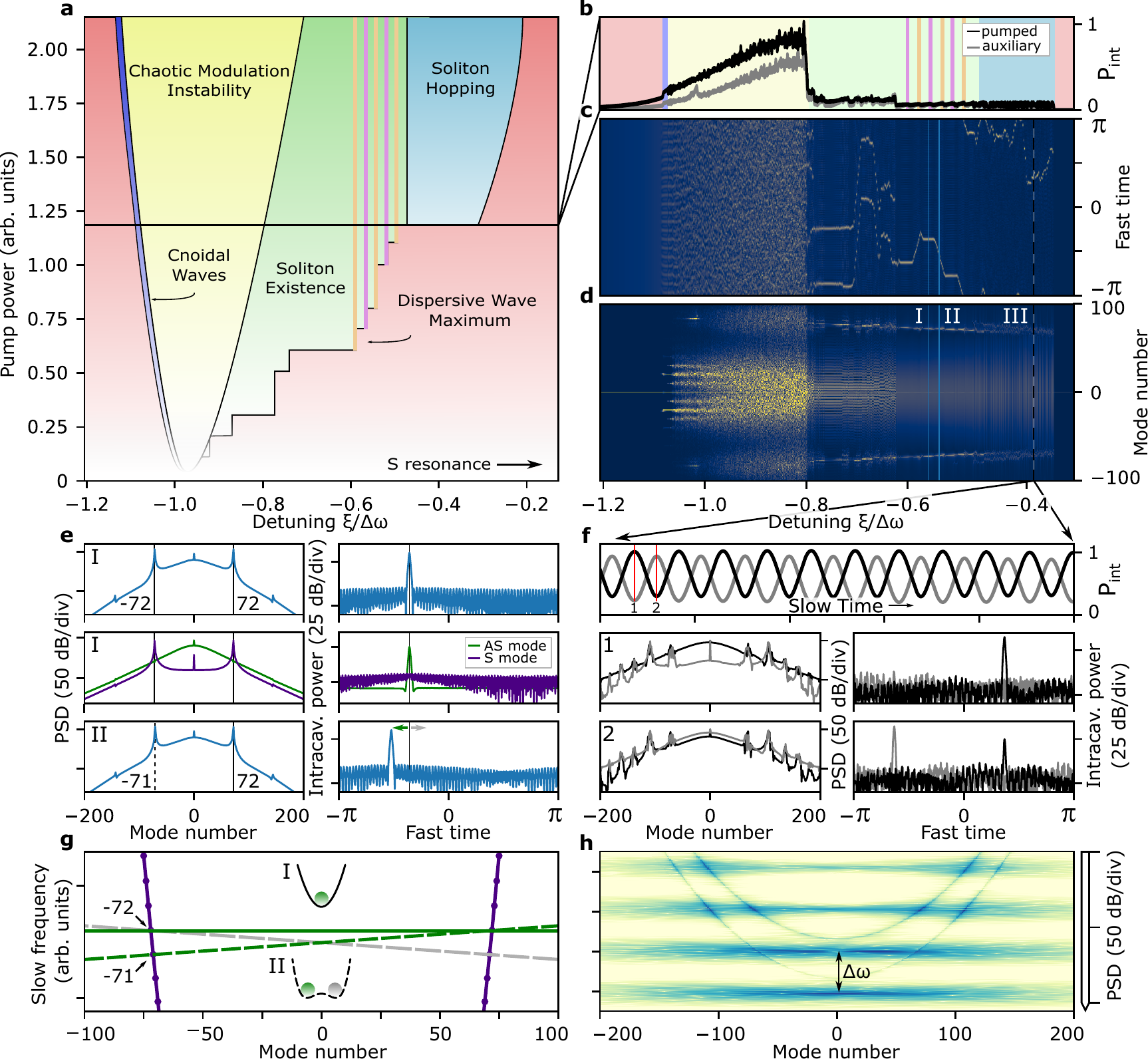}
	\caption{\textbf{Emergent dynamics revealed by numerical simulations of the photonic dimer.} \textbf{(a)} Schematic phase diagram showing system states at different values of pump power and laser detuning. A dark-blue area corresponds to the cnoidal waves (Turing rolls) emergence, yellow area - chaotic modulation instability, green area - stable GSs (Gear Solitons) existence, magenta and orange areas - maximum of DWs (Dispersive Waves) coexisting with GSs, blue area - soliton hopping regime. The red area depicts states in which only the continuous wave background is observed. The diagram is obtained by averaging over 10 realizations. \textbf{(b)} An example of the average intracavity power dependence on the laser detuning. Colours depicting different regions in diagram (a) are preserved. \textbf{(c)} Corresponding intracavity field and \textbf{(d)} power spectral density (PSD) evolution in the pumped resonator. \textbf{(e)} I and II show cross-sections of the plots (c) and (d) at the corresponding laser detuning values. I (bottom) shows the supermode decomposition of the state I. \textbf{(f)} Soliton hopping dynamics. The intracavity field is extracted at point III in plots (c-d) and propagated with a fixed laser detuning. Plots 1,2 show cases of maximum power in the pumped and auxiliary resonators, respectively. \textbf{(g-h)} Schematic representation of the nonlinear dispersion relation showing the symmetry breaking mechanism (g) and its numerical reconstruction for the soliton hopping state (h).}
	\label{fig:2}
\end{figure*}
Nonlinear dynamics in the photonic dimer can be described using a Hamiltonian formalism. We consider here a tight binding model with Kerr nonlinearity, such that the corresponding Hamiltonian can be written as:
\begin{linenomath*}
\begin{equation}
 \hat{H}_{\mathrm{d}} %d for dimer
   =  -\hbar J (\sum_{\mu} {\opa^{\dagger}_{1,\mu} }\opa_{2,\mu}+\opa^{\dagger}_{2,\mu} \opa_{1,\mu})  +  \hat{H}_1 + \hat{H}_2,
\label{eq:hamiltonian_dimer}
\end{equation}\end{linenomath*}
wherein $J$ denotes the evanescent coupling between the resonators. We note that in contrast to the Bose-Hubbard dimer case~\cite{Wim2017Quantum}, each resonator considered here has an infinite set of optical bosonic modes (e.g. $\opa_{1,\mu}$), designated by the mode index $\mu$ and distributed according to the microresonator dispersion relation $D_{\mathrm{int}}(\mu) = \omega_{\mu}-(\omega_{0}+\mu D_1)$ (cf.~Fig.~\ref{fig:1}d,e). As the resonators have almost identical free spectral ranges, the hybridization occurs among all considered modes of both resonators. In the DKS state the cavity dispersion and nonlinearity are exactly balanced for each mode $\mu$ such that the parabolic dispersion profile collapses into a straight line (cf.~Fig.~\ref{fig:1}d,e). $\hat{H}_{1,2}$ represent the conventional single resonator Hamiltonian that can be expressed as~\cite{Matsko2005Optical,Chembo2016Quantum}:
\begin{equation}
 \hat{H}_i = \hbar\sum_\mu \omega_{i,\mu} \opa^\dagger_{i,\mu}\opa_{i,\mu} - \hbar g_\mathrm{K}/2 ( \sum_{\mu} (\hat{a}_{i,\mu}  + \hat{a}_{i,\mu}^\dagger) )^4,
\label{eq:hamiltonian_single}
\end{equation} 
where the single photon Kerr shift is given by $g_\mathrm{K} = \frac{\hbar \omega_0^2 c n_2}{n_0^2 V_{\mathrm{eff}}}$, $c$ stands for the speed of light in vacuum, $V_{\mathrm{eff}}$ is the effective nonlinear mode volume, $n_0$ and $n_2$ are linear and nonlinear refractive indexes, respectively. Dissipation is introduced using the standard approach for open quantum systems, where we assume the regime that $J\gg\kappa$, where $\kappa$ denotes the cavity decay rate of an individual mode designated by its creation operator $\opa^\dagger_{1,\mu}$ and $\opa^\dagger_{2,\mu}$.

We next introduce the notion of orthogonal \textit{cavity supermodes} (see Fig.~\ref{fig:1}b). We refer to the supermodes with a higher frequency as antisymmetric (AS) and with a lower frequency as symmetric (S). The hybridized supermodes are given by:
\begin{linenomath*}
	\begin{align} 
	\opa_{\sym,\mu}& = \alpha\opa_{1,\mu} + \beta \opa_{2,\mu}\notag\\
	\opa_{\ant,\mu} & = \beta\opa_{1,\mu} - \alpha\opa_{2,\mu},
\label{eq:supmodes_q}
\end{align}\end{linenomath*} 
with coefficients $\alpha,\beta = \sqrt{1\mp d}/\sqrt{2},$ and the normalized inter-resonator detuning $d = \delta/\Delta\omega$ (which accounts for a detuning between the resonator modes, cf. Fig.~\ref{fig:1}b). The frequency splitting of the normalized modes is given as $\Delta\omega = \sqrt{4 J^2 + \delta^2}$. 

Introducing the change of variables Eq.~\ref{eq:supmodes_q} into Eq.~\ref{eq:hamiltonian_dimer}, we find a set of nonlinear coupling terms representing the interaction between the supermodes. 
With the notation $\opfwm^{\mu,\mu',\mu''}_{\sigma_1,\sigma_2,\sigma_3,\sigma_4} = \OPC{a}_{\sigma_1,\mu}\OPC{a}_{\sigma_2,\mu'}\OP{a}_{\sigma_3,\mu''}\OP{a}_{\sigma_4,\mu+\mu'-\mu''}$; $\sigma_i$ = \{a,s\}, where a and s stand for antisymmetric and symmetric mode, the resulting Hamiltonian (see Supplementary Information section S1 for details) can be expressed as follows:

\begin{align}
&\OPH =\notag \\ 
&\hbar\sum_{\mu}\left[\omega_\mu( \OPCaas[,\mu]\OPaas[,\mu] +\OPCas[,\mu]\OPas[,\mu])+\frac{\Delta\omega}{2}( \OPCaas[,\mu]\OPaas[,\mu] -\OPCas[,\mu]\OPas[,\mu])\right] \notag\\
& - \hbar\frac{g_\mathrm{K}}{2}\sum_{\mu,\mu',\mu''}\left[\frac{1}{2}(1+d^2)\underbrace{(\opfwm^{\mu,\mu'\!,\mu''\!}_{\sym{,}\sym{,}\sym{,}\sym} + \opfwm^{\mu,\mu'\!\!,\mu''}_{\asym,\asym,\asym,\asym})}_{\mathrm{even\, and \, intra-band}}\right.\notag\\
& -d\sqrt{1-d^2}\underbrace{ (\opfwm^{\mu,\mu'\!,\mu''}_{\,\sym,\sym,\sym,\asym}+ \opfwm_{\sym,\asym,\sym,\sym}^{\mu,\mu'\!\!,\mu''\!} - \opfwm^{\mu,\mu'\!,\mu''}_{\,\sym,\asym,\asym,\asym} - \opfwm^{\mu,\mu'\!,\mu''}_{\,\asym,\asym,\sym,\asym})}_{\mathrm{odd\, and \, inter-band}}\notag\\
&\left.  + \frac{1}{2}(1-d^2)\underbrace{  (\opfwm^{\mu,\mu'\!,\mu''}_{\,\sym,\sym,\asym,\asym} + 4 \opfwm^{\mu,\mu'\!,\mu''}_{\,\sym,\asym,\sym,\asym} + \opfwm^{\mu,\mu'\!,\mu''}_{\,\asym,\asym,\sym,\sym})}_{\mathrm{even\, and \, inter-band}}\right].
\label{eq:hamiltonian_hybrid}
\end{align}

GSs can be generated in both supermodes, however, the modification of the spectral profile has been observed only in the AS case, i.e. by tuning over the upper parabola on the dispersion relation. Indeed, the phase-matching conditions can be fulfilled when the solitonic line crosses the lower (S supermodes) parabola which creates \textit{novel Four-Wave Mixing (FWM) pathways between AS and S supermodes}. The presence of such phase-matched interactions makes nonlinear dynamics of the photonic dimer remarkably more rich and complex than in the single resonator case.
Novel nonlinear processes can be divided into two categories according to the number of photons from each of the supermodes involved: even and odd. Remarkably, even and odd processes have different efficiencies depending on the normalized inter-resonator detuning as follows from Eq.~\ref{eq:hamiltonian_hybrid}.

The GS is maintained in the AS supermodes by the cascaded FWM processes similar to the single resonator case (see Fig.~\ref{fig:1}d) represented by the term $\opfwm^{\mu,\mu'\!\!,\mu''}_{\asym,\asym,\asym,\asym}$ in Eq.~\ref{eq:hamiltonian_hybrid}. Since the photons stay within the same supermode family, we refer to this process as \emph{ even and intra-band}. \emph{Inter-band} FWM processes are shown in Fig.~\ref{fig:1}e. Process \#2 is odd since two photons annihilated in the AS supermode create two photons: one in the AS and one in the S supermodes, while process \#3 is even. They are represented by $\opfwm^{\mu,\mu'\!,\mu''}_{\,\sym,\asym,\asym,\asym}$ and $\opfwm^{\mu,\mu'\!,\mu''}_{\,\sym,\sym,\asym,\asym}$ in Eq.~\ref{eq:hamiltonian_hybrid}, respectively. Inter-band FWM generates DW when the solitonic line is in the vicinity of an S supermode. DWs are represented by two symmetrically-spaced Lorentzian profiles in the S supermode spectrum. Maxima of the Lorentzians occur exactly at the mode where the crossing occurs. Interference between the hyperbolic secant (in AS supermodes) and Lorentzians (in S supermodes) results in a Fano-shaped spectrum as shown in Fig.~\ref{fig:1}c. Remarkably, a similar spectral feature has been discovered numerically in a $\mathcal{PT}$-symmetric dimer~\cite{milian2018cavity}, where the appearance of the Fano strictures was interpreted as the result of modulation instability.

\begin{figure*}
	\centering
	\includegraphics[width=\linewidth]{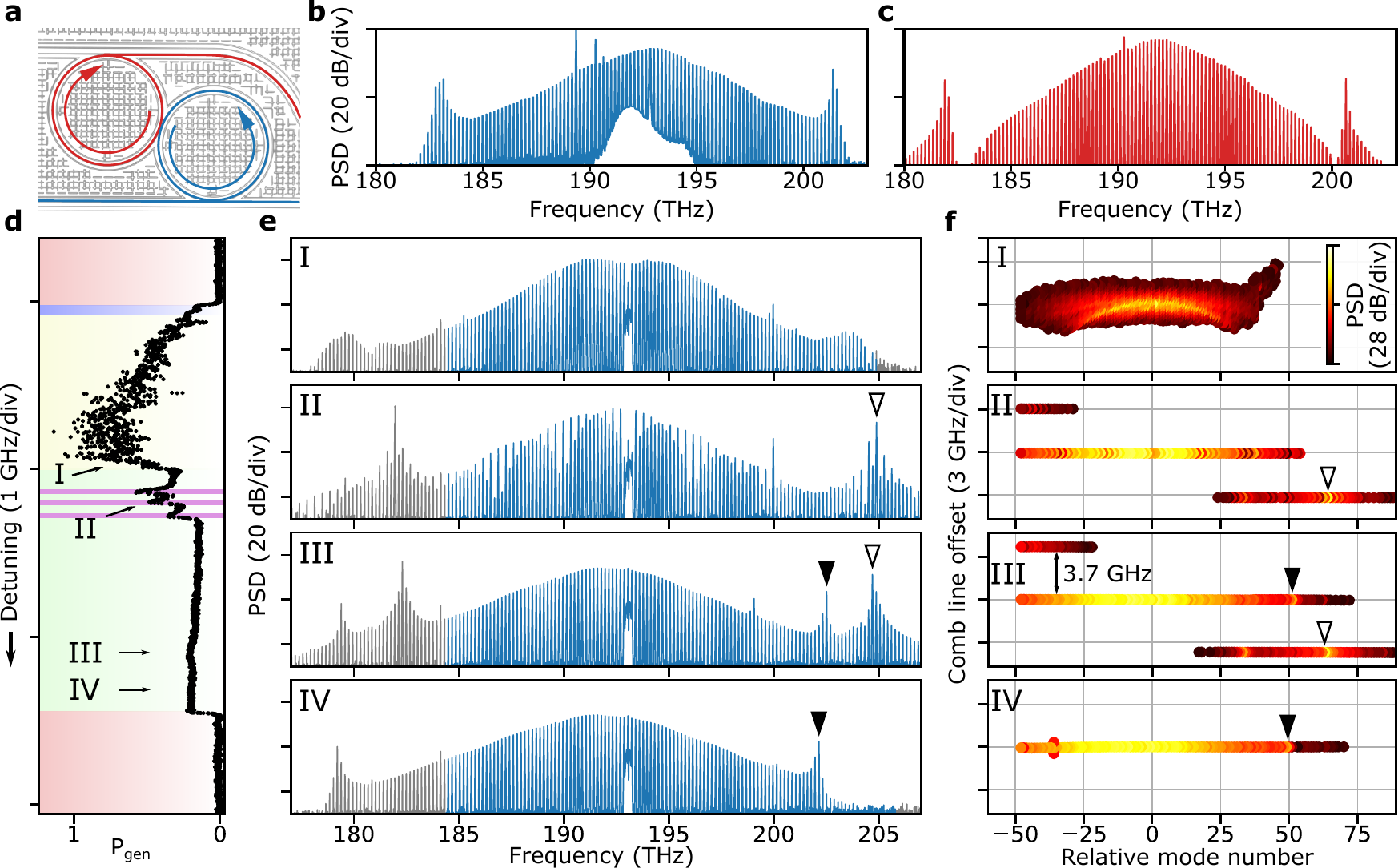}
	\caption{\textbf{Kerr comb reconstruction of photonic dimer states}
	\textbf{(a)} Microscope image of coupled Si$_3$N$_4$ resonators. Arrows indicate the propagation direction of the soliton in the resonators. \textbf{(b,c)} GS (Gear Soliton) spectra recorded at the bus (blue) and drop (red) waveguides.
	\textbf{(d)} Power of generated light while the pump laser is scanned across the antisymmetric resonance. Colors correspond to Fig.~\ref{fig:2}b. \textbf{(e, I-IV)} Optical spectra of generated light at different pump laser detuning. I - modulation instability, II - three GSs state, III  - stable GS with offset DW (Dispersive Wave), IV - stable GS with commensurate DW. Blue coloring indicates region of comb reconstruction. \textbf{(f, I-IV)} Kerr comb reconstruction of different states showing the frequency offsets of the comb lines versus their relative mode numbers (lighter colours correspond to higher comb line power). I - modulation instability follows the parabolic dispersion profile, II, III - resonant DW appears with carrier-envelope-offset from main soliton comb. IV - GS line with commensurate DW signature.}
	\label{fig:3}
\end{figure*}

\section{Exploring the phase diagram of the photonic dimer}

Even though weakly coupled \cite{Jang2018Synchronization} and size-mismatched \cite{Xue2019Super,Jang2018Synchronization} microring resonators have been recently studied in the context of DKSs synchronization and enhancement of the nonlinear conversion efficiency, the deployment of large nonlinear photonic lattices requires an understanding of dynamics of strongly coupled and uniform systems of nonlinear cavities.

In order to investigate the phase diagram of the strongly coupled nonlinear photonic dimer, we perform numerical simulations of the coupled-mode equations realizing a method proposed in~\cite{Hansson2014On} (see Supplementary Information section S2 for details). Fig.~\ref{fig:2}a shows a schematic phase diagram. The initial dynamics is found to be similar to the single resonator case~\cite{Godey2014Stability}. We observe the formation of primary combs followed by the cnoidal waves (Turing rolls)~\cite{qi2019dissipative} in both resonators (dark blue area in Fig.~\ref{fig:2}a,b). 
Fig.~\ref{fig:2}c,d provide the underlying evolution of intracavity power (spatiotemporal diagram) and power spectral density of the pumped resonator. The appearance of the spectral components enhanced due to the supermode interaction can be seen as converging distinct lines in Fig.~\ref{fig:2}d.

Significant divergence from the single resonator case is observed in the soliton existence range. Stable solitons are strongly perturbed by the periodic emergence of DWs resulting from the inter-band FWM interactions. The soliton existence range is represented by the green area in Fig.~\ref{fig:2}a. The DW maxima are shown by vertical magenta and orange lines. The interaction between supermodes leads to the periodic increase of the average intracavity power in both resonators (see Fig.~\ref{fig:2}b). High amplitude DWs perturb the GS state which can lead to its decay or decreasing of the soliton number. The strength of this effect directly depends on the number of solitons and, therefore, the system is naturally forced towards the single GS state where the perturbation is the weakest~\cite{bao2017spatial}. Since the values of the laser detuning at which the resonant enhancement of the DW occurs are related to the discretization of the lower parabola, the soliton steps are discretized as well according to the position of the resonances. The phase diagram is averaged over 10 realizations and the values of the laser detuning at which the single soliton state decays most probably are plotted.

The difference between the two DWs maxima (magenta and orange) can be seen in Fig.~\ref{fig:2}c. Maxima depicted in magenta correspond to the cases when the solitonic line on the dispersion relation crosses exactly points on the lower parabola (Fig.~\ref{fig:2}g, solid line). The resonant interaction becomes efficient which leads to the increasing average intracavity power~\cite{yi2017single}. A further change of the pump laser detuning brings GS between two discrete points on the S supermodes parabola where it demonstrates an unstable behaviour. This instability leads to a \textit{symmetry breaking} during which the GS acquires a group velocity (see Supplementary Video~1). The sign of the group velocity varies arbitrarily from one realization to another. The additional group velocity corresponds to a tilt of the solitonic line on the nonlinear dispersion relation~\cite{Leisman2019Effective} (Fig.~\ref{fig:2}g, green and gray dashed lines). We found out that the group velocity self adjusts to a value, which precisely ensures that the dispersionless soliton line connects again with discrete points on the S parabola. Surprisingly, intermediate values are not chosen. Fulfillment of the resonance condition, as previously, leads to the increase of average intracavity power which is depicted by orange lines in Fig.~\ref{fig:2}a,b. The optical spectrum and intracavity power profiles of two different maxima of the DWs amplitude are shown in Fig.~\ref{fig:2}e (I and II). The change of the group velocity of the GS leads to an asymmetry in the wings of the spectrum and a displacement of Fano-shaped peaks by one mode number (dashed line in Fig.~\ref{fig:2}e~II). Decomposition into the supermodes according to Eq.~\ref{eq:supmodes_q} demonstrates a clear separation of the GS and DWs belonging to different supermodes (Fig.~\ref{fig:2}e~I (bottom)).

\begin{figure*}
	\centering
	\includegraphics[width=\linewidth]{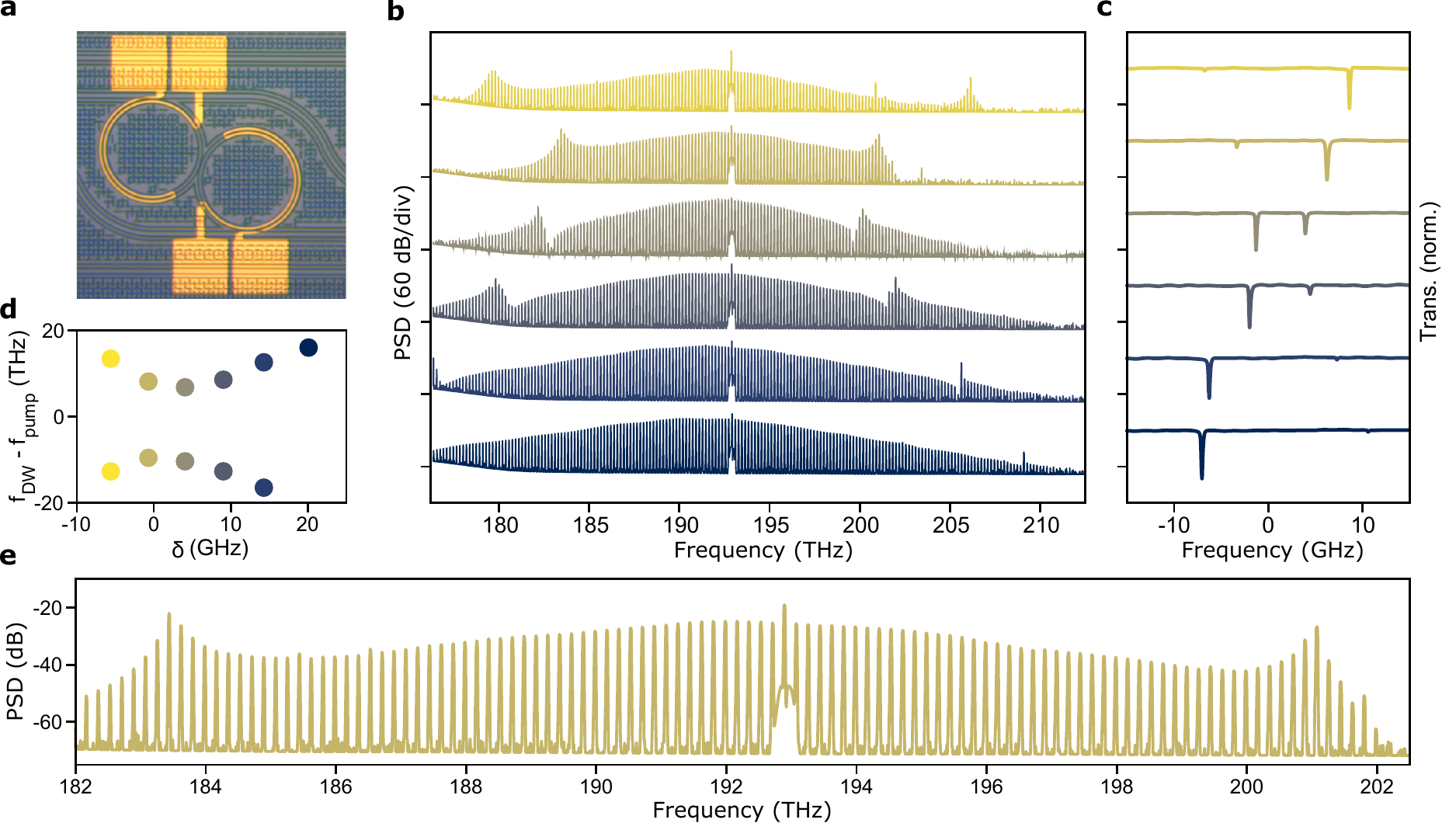}
	\caption{\textbf{Electrical control of gear solitons and dispersive wave interference.} \textbf{(a)} Microscope image of a chip containing two coupled Si$_3$N$_4$ resonators partially covered by gold microheaters. \textbf{(b)} Optical spectra recorded at the drop waveguide at different inter-resonator detunings $\delta$ (color code refers to panel d). \textbf{(c)} Linear measurements of the transmission. \textbf{(d)} Position of the spectral maxima associated with the DWs as a function of the inter-resonator detuning $\delta$. \textbf{(e)} Zoom of the second spectrum from the top in plot (b) showing that the energy of the comb lines remains similar over a wide range of frequencies. Colours in all plots are preserved.}
	\label{fig:4}
\end{figure*}

A novel effect is found at higher values of the pump power. After reaching a critical value of the laser detuning, the system exhibits periodic energy exchange between the coupled resonators while being in the GS state. This dynamical regime is referred to as \textit{soliton hopping} and is indicated by the blue area in Fig.~\ref{fig:2}a,b. This process is similar to the so-called self-pulsation effect discovered in single-mode dimers~\cite{grigoriev2011resonant,abdollahi2014analysis}. However, in the multimode case, the presence of solitons allows for the synchronous pulsation of a large number of longitudinal modes which enables the hopping of the spatiotemporal structure as a whole, emphasizing the quasi-particle nature of the DKSs. The soliton hopping regime is studied by extracting the complex amplitude from both resonators at point III in Fig.~\ref{fig:2}c,d and numerically propagating it with a fixed laser detuning (see Fig.~\ref{fig:2}f). The average power in the coupled resonators oscillates in counter-phase. In this regime, the GS is strongly projected onto the S supermode which leads to periodic constructive and destructive interference. The soliton hopping frequency coincides well with the gap between the hybridized parabolas in the dispersion relation. Power spectral densities and intracavity powers at maximum power in the pumped resonator and maximum power in the auxiliary resonator are shown in Fig.~\ref{fig:2}f (1,2), respectively. Additional sidebands often associated with breathing~\cite{Lucas2017Breathing} are observed in the optical spectra. The distance between sidebands decreases with increasing mode number. This is similar to the case of Kelly sidebands which appear during periodic perturbation of a solitonic state~\cite{kelly1992characteristic}. The origin of the sidebands can be easily understood by analysing the corresponding nonlinear dispersion relation (Fig.~\ref{fig:2}h). We numerically reconstruct the nonlinear dispersion relation by taking a double Fourier transform of the spatiotemporal diagram. The periodic soliton hopping here is represented by a set of equally spaced lines (ladder) separated by $\Delta \omega$, which corresponds to the hopping frequency. The sidebands appear exactly at the points were the ladder crosses the S and AS parabolas.

\section{Experimental results} % <<-- Johann 
\subsection{Experimental observation of Gear Solitons on the Si$_3$N$_4$ platform}

Experimental investigation of the phenomena described above requires microresonators with anomalous dispersion, ultra-low loss and exceptional uniformity, which are fabricated with the photonic Damascene reflow process on Si$_3$N$_4$~\cite{Pfeiffer2018Ultra,liu2018ultralow}. The DKSs were generated in a pair of 181~GHz resonators (cf.~Fig.~\ref{fig:3}a) with a frequency-dependent evanescent coupling of $J/2\pi$~=~3-6~GHz having intrinsic and external coupling loss rates $\kappa_\textrm{0},\kappa_\textrm{ex} \approx $~$2\pi \cdot$ 25~MHz, by laser tuning~\cite{Herr2014Temporal} (cf.~Fig.~\ref{fig:3}b,c). Further details of the experimental setup are described in the Supplementary Information sections S3-S5.

Fig.~\ref{fig:3}d depicts the power of the light generated in the dimer as the pump laser frequency is reduced, crossing the AS resonance. Similar to the single resonator case, a chaotic modulation instability state is observed, which collapses to a low-noise GS state as the laser crosses the resonance from blue to red detuning. At low pump power (200~mW in the waveguide), a manifold of multisoliton GS can be accessed (cf.~Fig.~\ref{fig:3}e,f II). At high pump power (900~mW in the waveguide), the multisoliton states rapidly collapse to a single GS state due to their mutual disruption in the presence of strong resonant DW formation in the S supermode~\cite{bao2017spatial}. We observe two types of dual DWs in hybridized supermodes: \emph{Fano} shaped and \emph{Lorentzian} shaped. Fano-shaped DWs are observed continuously during the tuning scan, while Lorentz-shaped DWs appear resonantly at a fixed detuning (cf.~Fig.~\ref{fig:3}e III). We investigate the coherence among the DW and the soliton using absolute frequency measurements of each frequency component of the spectrum~\cite{herr2012universal}. 

We find that the Fano-shaped hybrid DW is bound to the soliton, i.e. it has identical repetition rate $f_\mathrm{rep}$ and  carrier-envelope-offset frequency $f_\mathrm{ceo}$ and constitute a \emph{commensurate} continuation of the primary soliton frequency comb, facilitating their exploitation for metrology \cite{brasch2017self,spencer2018optical}. In contrast to the single-particle LLE intuition, the Lorentz shaped hybrid DW is actually generated at a different $f_\mathrm{ceo}$ (cf.~Fig.~\ref{fig:3}f III) and hence constitutes an \emph{incommensurate} secondary frequency comb. Numerical studies (see Supplementary Information section S2) reveal that the cause of the emergence of incommensurate DWs is the frequency-dependency of the coupling rate $J$ inherent to the evanescent coupling scheme. Strikingly the spectrum still reveals multisoliton interference patterns\cite{brasch2016photonic} - an indication of a fully coherent comb in the single-particle LLE - which, however, in the dimer case, does not always lead to a fully coherent optical comb (Fig.~\ref{fig:3}e, panel II). It is noteworthy that the DWs continue the spectral line pattern of multisoliton and even soliton crystal states (cf.~SI Fig.~8). The observation of commensurate and incommensurate dispersive waves in the photonic dimer motivates a deeper analysis of Eq. \ref{eq:hamiltonian_dimer}. The prefactors of even and odd inter-band FWM transitions reveal that for vanishing normalized detuning $d = 0$, the odd inter-band transition, which inherently preserves the frequency offset of the soliton in the DW, vanishes as well. Strikingly and counter-intuitively, our numerical studies show that a degree of asymmetry either in dissipation (i.e. asymmetric external coupling $\kappa_{\mathrm{ex}}/2\pi$) or a finite inter-resonator detuning $\delta$ is required for the formation of fully coherent dissipative structures in the S modes.

\subsection{Electrical control of hybrid dispersive waves} 
Having studied emergent nonlinear phenomena in the nonlinear dimer, we show some practical relevance of the additional control provided by this system. Since the Fano shape in the optical spectrum originates from the interference of the soliton in the AS supermode family and dispersive waves in the S supermode family, its position is determined by the intersection of the solitonic line with the S parabola given by $ \mu_{DW} = \sqrt{\frac{2(\Delta\omega-\zeta)}{D_{2}}}$. Therefore, the position of the enhanced spectral lines can be readily tuned by thermally varying the inter-resonator detuning $\delta$~\cite{miller2015tunable}. We suppress the appearance of incommensurate DWs by operating the device with strongly asymmetric external coupling. Thermal tuning up to 19~GHz is achieved via integrated gold microheaters (see Fig.~\ref{fig:4}a.), which corresponds to the tuning of DW positions over a range of 10~THz. While Fano-shaped \cite{Guo2017Intermode} and single-mode DWs \cite{yi2017single} have been observed before, the photonic dimer case presents the opportunity to electrically control the Fano lineshape, i.e. the spectral regions where constructive interference takes place. In particular, when the value of $\delta$ is negative, the Fano shapes are oriented such that the resulting solitonic spectra are flat over a broad spectral range. This unique feature, together with the demonstrated tunability makes the soliton dimer particularly suitable for telecommunication-oriented applications~\cite{marin2017microresonator}.

\section{Conclusion}
In conclusion, studying linearly coupled driven nonlinear microresonators, we have observed emergent nonlinear dynamics, i.e. dynamics that is not contained in the single-particle LLE. The observations range from novel resonant Fano dispersive waves and soliton hopping to the surprising relation between the phase coherence of dissipative structures and imperfect mode hybridization. These observations cannot be predicted nor anticipated from the single-particle LLE, and indicate the rich nonlinear dynamics that can occur when increasing the system size further to larger lattices. Thereby, the current work represents the first step in the development of integrated soliton lattices. Experimental and theoretical approaches demonstrated here, including the representation of the hybridized dispersion relation, spectral analysis of the FWM pathways, and thermal tuning of the Fano resonances, can serve as a foundation to further studies of a large one- and higher-dimensional soliton lattices.  Our study underscores Anderson's statement 'more is different' \cite{anderson1972more} and extends it to driven dissipative nonlinear cavities, which due to recent progress in photonic materials and fabrication, can now be arranged in extended lattices. 
The latter are of strong current interest in frameworks of nonlinear all-optical computing~\cite{McMahon2016fully} and neural network implementations~\cite{Shen2018Deep}, topological photonics~\cite{ozawa2019topological2}, digitally programmable chromatic dispersion systems~\cite{Yao2018Gate}, and $\mathcal{PT}$-symmetric systems~\cite{hodaei2014parity}. Finally, we anticipate the possibility to create global Bloch solitons in a lattice of coupled resonators containing DKSs in each of them.

\section{Acknowledgments}
The authors would like to thank A. Tusnin and M. Karpov for fruitful discussions. This publication was supported by Contract 18AC00032 (DRINQS) from the Defense Advanced Research Projects Agency (DARPA), Defense Sciences Office (DSO). This work was further supported by the European Union’s Horizon 2020 Program for Research and Innovation under grant no. 846737 (Marie Skłodowska-Curie IF CoSiLiS), 812818 (Marie Skłodowska-Curie ETN MICROCOMB), 722923 (Marie Skłodowska-Curie ETN OMT), and 732894 (FET Proactive HOT). Si$_3$N$_4$ samples were fabricated and grown in the Center of MicroNanoTechnology (CMi) at EPFL. Microheater integration was performed at the Binnig and Rohrer Nanotechnology Center at IBM Research Europe.
\bibliography{citations}

\end{document}

\newcommand{\abs}[1]{\left|#1\right|}
\newcommand{\abssq}[1]{\abs{#1}^2}%norm squared
\newcommand{\diff}[1]{\mathrm{d}#1}
\newcommand{\drt}{\tfrac{\mathrm{d}}{\diff{t}}}
\renewcommand{\Re}{\text{Re}}%real part
\renewcommand{\Im}{\text{Im}}%real part

\newcommand{\AS}{A_\mathrm{s}}
\newcommand{\AAS}{A_\mathrm{as}}
\newcommand{\ASmu}{A_{\mathrm{s},\mu}}
\newcommand{\AASmu}{A_{\mathrm{as},\mu}}

%shortcuts for hamiltonian calculation
\newcommand{\OP}[1]{\hat{#1}} %operators
\newcommand{\OPC}[1]{\hat{#1}^\dagger} % adjoint
\newcommand{\OPH}{\OP{H}}
\newcommand{\OPa}{\OP{a}}
\newcommand{\OPCa}{\OPC{a}}
\newcommand{\OPb}{\OP{b}}
\newcommand{\OPCb}{\OPC{b}}
\newcommand{\Jc}{J}
\newcommand{\gc}{g_\mathrm{K}}
\newcommand{\asym}{\mathrm{a}}
\newcommand{\OPas}[1][]{\OP{a}_{\sym#1}}
\newcommand{\OPaas}[1][]{\OP{a}_{\asym#1}}
\newcommand{\OPCas}[1][]{\OPC{a}_{\sym#1}}
\newcommand{\OPCaas}[1][]{\OPC{a}_{\asym#1}}
\newcommand{\opfwm}{\OP{k}}
	
%\title{Coupled Microcavity Soliton Networks}
%\title{Dissipative Kerr Soliton Lattices}
%\title{Kerr Solitons in Photonic Lattices}
%\title{Dimerized Disspative Kerr Solitons}
%\title{Dissipative Kerr solitons resonator dimers}
\title {Supplementary information: Emergent Nonlinear Phenomena in a Driven Dissipative Photonic Dimer}

\author{A.~Tikan$^1$, J.~Riemensberger$^1$, K.~Komagata$^{1,3}$, S.~H\"onl$^2$ ,M.~Churaev$^1$, C.~Skehan$^1$,  H.~Guo$^{1,4}$, R.~N.~Wang$^1$, J.~Liu$^1$, P.~Seidler$^2$, T.J.~Kippenberg}
\affiliation{Institute of Physics, Swiss Federal Institute of Technology Lausanne (EPFL), CH-1015 Lausanne, Switzerland \\ $^2$IBM Research Europe, Säumerstrasse 4, CH-8803 Rüschlikon, Switzerland \\ $^3$ Present address: Laboratoire Temps-Fréquence, Institut de Physique, Université de Neuchâtel, CH-2000 Neuchâtel, Switzerland \\
$^4$ Present address: Key Laboratory of Specialty Fiber Optics and Optical Access Networks, Shanghai University, 200444 Shanghai, China}
\date{\today}
\pacs{}
\maketitle
\section{Derivation of the quantum Hamiltonian in conserved system}
We start with the Hamiltonian of two non-linear (coefficient $\gc$) bosonic modes $\OPa,\OPb$ detuned by $\delta$ and linearly coupled with strength $\Jc$:
\begin{equation}
\OPH = \hbar\omega_0 ( \OPCa\OPa+\OPCb\OPb) + \hbar\frac{\delta}{2} ( \OPCa\OPa-\OPCb\OPb) - \hbar\Jc(\OPCa\OPb+\OPCb\OPa) - \hbar\frac{\gc}{2}( \OPCa\OPCa\OPa\OPa+\OPCb\OPCb\OPb\OPb).
\label{si_eq:1}
\end{equation}
Defining the antisymmetric and symmetric supermodes which diagonalize the linear part of the system as follows:
\begin{align}
\OPas &= \alpha \OPa + \beta \OPb & \OPaas &= \beta \OPa - \alpha \OPb\\
\OPCas &= \alpha \OPCa + \beta \OPCb & \OPCaas &= \beta \OPCa - \alpha \OPCb\\
\alpha &= \frac{\sqrt{1-d}}{\sqrt{2}} & \beta &= \frac{\sqrt{1+d}}{\sqrt{2}},
\label{si_eq:4}
\end{align}
with $d = \delta/\Delta\omega$, $\Delta\omega = \sqrt{\delta^2+4\Jc^2}$. We can verify the commutator relation:
\begin{equation}
[\OPas,\OPCas] = \alpha^2[\OPa,\OPCa] + \beta^2[\OPb,\OPCb] = \frac{1-d}{2}+\frac{1+d}{2} = 1.
\end{equation}
Using the definition of the hybridized modes we can state that:
\begin{align}
\OPa &=\alpha \OPas + \beta\OPaas& \OPb &=\beta\OPas - \alpha \OPaas.
\end{align}
Therefore, we can obtain the following relations:
\begin{align}
\OPCa\OPa &= \alpha^2\OPCas\OPas + \beta^2\OPCaas\OPaas + \alpha\beta(\OPCas\OPaas+\OPCaas\OPas)\\
\OPCb\OPb &= \beta^2\OPCas\OPas + \alpha^2\OPCaas\OPaas - \alpha\beta(\OPCas\OPaas+\OPCaas\OPas)\\
\OPCa \OPb &= \alpha\beta\OPCas\OPas -\alpha\beta \OPCaas\OPaas -\alpha^2\OPCas\OPaas + \beta^2 \OPCaas\OPas\\
\OPCb \OPa &= \alpha\beta\OPCas\OPas -\alpha\beta \OPCaas\OPaas +\beta^2\OPCas\OPaas -\alpha^2 \OPCaas\OPas.
\end{align}
The same can be done with nonlinear terms of the Hamiltonian:
\begin{align}
\OPCa\OPCa\OPa\OPa & = (\alpha^2\OPCas \OPCas + \beta^2 \OPCaas\OPCaas +2\alpha\beta \OPCas\OPCaas) (\alpha^2\OPas \OPas + \beta^2 \OPaas\OPaas +2\alpha\beta \OPas\OPaas)\notag\\
&= \alpha^4\OPCas\OPCas\OPas\OPas + 2\alpha^3\beta\OPCas\OPCas\OPas\OPaas + \alpha^2\beta^2\OPCas\OPCas\OPaas\OPaas\notag \\
&\quad+2\alpha^3\beta \OPCas\OPCaas\OPas\OPas + 4 \alpha^2\beta^2\OPCas\OPCaas\OPas\OPaas + 2\alpha\beta^3\OPCas\OPCaas\OPaas\OPaas\notag\\
&\quad + \alpha^2\beta^2\OPCaas\OPCaas \OPas\OPas + 2\alpha\beta^3\OPCaas\OPCaas\OPas\OPaas + \beta^4 \OPCaas\OPCaas\OPaas\OPaas\\
\OPCb\OPCb\OPb\OPb & = \beta^4\OPCas\OPCas\OPas\OPas - 2\alpha\beta^3\OPCas\OPCas\OPas\OPaas + \alpha^2\beta^2\OPCas\OPCas\OPaas\OPaas\notag \\
&\quad-2\alpha\beta^3 \OPCas\OPCaas\OPas\OPas + 4 \alpha^2\beta^2\OPCas\OPCaas\OPas\OPaas -2\alpha^3\beta\OPCas\OPCaas\OPaas\OPaas\notag\\
&\quad + \alpha^2\beta^2\OPCaas\OPCaas \OPas\OPas - 2\alpha^3\beta\OPCaas\OPCaas\OPas\OPaas + \alpha^4 \OPCaas\OPCaas\OPaas\OPaas.
\end{align}
From expression~\ref{si_eq:4} one can deduce that:
\begin{align}
 (\beta^2 - \alpha^2) &= \frac{\delta}{\Delta\omega} & \alpha\beta &= \frac{\Jc}{\Delta\omega}.
\end{align}
The resulting Hamiltonian is given by:
\begin{align}
\OPH/\hbar &= \omega_0 ( \OPCaas\OPaas +\OPCas\OPas)+ \frac{\Delta\omega}{2}( \OPCaas\OPaas -\OPCas\OPas) - \frac{\gc}{2}\left[(\alpha^4 + \beta^4)(\OPCas\OPCas\OPas\OPas + \OPCaas\OPCaas\OPaas\OPaas)\right.\notag\\
&\quad+ 2\alpha\beta(\alpha^2-\beta^2)(\OPCas\OPCas\OPas\OPaas + \OPCas\OPCaas\OPas\OPas - \OPCas\OPCaas\OPaas\OPaas - \OPCaas\OPCaas\OPas\OPaas)\notag\\
&\left. \quad + 2\alpha^2\beta^2 (\OPCas\OPCas\OPaas\OPaas + 4 \OPCas\OPCaas\OPas\OPaas + \OPCaas\OPCaas\OPas\OPas)\right]\\
&=\omega_0 ( \OPCaas\OPaas +\OPCas\OPas)+\frac{\Delta\omega}{2}( \OPCaas\OPaas -\OPCas\OPas) - \frac{\gc}{2}\left[\frac{1}{2}(1+d^2)(\OPCas\OPCas\OPas\OPas + \OPCaas\OPCaas\OPaas\OPaas)\right.\notag\\
&\quad-d\sqrt{1-d^2}(\OPCas\OPCas\OPas\OPaas + \OPCas\OPCaas\OPas\OPas - \OPCas\OPCaas\OPaas\OPaas - \OPCaas\OPCaas\OPas\OPaas)\notag\\
&\left. \quad + \frac{1}{2}(1-d^2) (\OPCas\OPCas\OPaas\OPaas + 4 \OPCas\OPCaas\OPas\OPaas + \OPCaas\OPCaas\OPas\OPas)\right]
\\
&=\omega_0 ( \OPCaas\OPaas +\OPCas\OPas) + \frac{\Delta\omega}{2}( \OPCaas\OPaas -\OPCas\OPas) - \frac{\gc}{2}\left[\frac{1}{2}(1+(\frac{\delta}{\Delta\omega})^2)(\OPCas\OPCas\OPas\OPas + \OPCaas\OPCaas\OPaas\OPaas)\right.\notag\\
&\quad-2\frac{\delta\Jc}{(\Delta\omega)^2}(\OPCas\OPCas\OPas\OPaas + \OPCas\OPCaas\OPas\OPas - \OPCas\OPCaas\OPaas\OPaas - \OPCaas\OPCaas\OPas\OPaas)\notag\\
&\left. \quad + 2 (\frac{\Jc}{\Delta\omega})^2 (\OPCas\OPCas\OPaas\OPaas + 4 \OPCas\OPCaas\OPas\OPaas + \OPCaas\OPCaas\OPas\OPas)\right].
\end{align}

Let us extend the Hamiltonian~\ref{si_eq:1} to the multimode case. We define the set of bosonic modes $\OPa_\mu, \OPb_\mu$, so the following expression can be obtained:
\begin{align}
\OPH = \hbar\sum_{\mu}\left[\omega_\mu\left(\OPCa_\mu \OPa_\mu + \OPCb_\mu \OPb_\mu\right) + \frac{\delta}{2}\left(\OPCa_\mu \OPa_\mu - \OPCb_\mu \OPb_\mu\right) - \Jc \left(\OPCb_\mu\OPa_\mu + \OPCa_\mu \OPb_\mu \right) \right]+ \OPH_{\mathrm{Kerr}}.
\end{align}
The last term defines the on-site Kerr nonlinearity. It is defined as $\OPH_{_\mathrm{Kerr}} = -\hbar\frac{\gc}{2}:\left(\sum_{\mu} (\OP{o}_\mu + \OPC{o}_\mu)\right)^4:$ for each site $\OP{o} = \OPa,\OPb$, where $:...:$ means the normal ordering of the operator \cite{Matsko2005Optical}. Furthermore, the rotating wave approximation is used, together with assumptions that the resonant frequencies are close to frequency grid, i.e. $\omega_\mu \approx \omega_0 + \mu D_1$. The Hamiltonian is then:
\begin{align}
\OPH &= \hbar\sum_{\mu}\left[\omega_\mu\left(\OPCa_\mu \OPa_\mu + \OPCb_\mu \OPb_\mu\right) + \frac{\delta}{2}\left(\OPCa_\mu \OPa_\mu - \OPCb_\mu \OPb_\mu\right) - \Jc \left(\OPCb_\mu\OPa_\mu + \OPCa_\mu \OPb_\mu \right) \right] \notag\\
&\quad- \hbar\frac{\gc}{2}\sum_{\mu,\mu',\mu''}\left(\OPCa_\mu\OPCa_{\mu'}\OPa_{\mu''}\OPa_{\mu+\mu'-\mu''} +\OPCb_\mu\OPCb_{\mu'}\OPb_{\mu''}\OPb_{\mu+\mu'-\mu''}\right).
\end{align}
By performing a similar transformation for each longitudinal mode, i.e. $\OPas[,\mu] = \alpha\OPa_\mu + \beta \OPb_\mu,~\OPaas[,\mu] = \beta\OPa_\mu -\alpha \OPb_\mu$, the linear part is diagonalized for for each pair of modes with index $\mu$. With the notation $\opfwm^{\mu,\mu',\mu''}_{\sigma_1,\sigma_2,\sigma_3,\sigma_4} = \OPC{a}_{\sigma_1,\mu}\OPC{a}_{\sigma_2,\mu'}\OP{a}_{\sigma_3,\mu''}\OP{a}_{\sigma_4,\mu+\mu'-\mu''}$, the Hamiltonian is expressed as
\begin{align}
\OPH = \,
&
\hbar\sum_{\mu}\left[\omega_\mu( \OPCaas[,\mu]\OPaas[,\mu] +\OPCas[,\mu]\OPas[,\mu])+\frac{\Delta\omega}{2}( \OPCaas[,\mu]\OPaas[,\mu] -\OPCas[,\mu]\OPas[,\mu])\right] \notag\\
& - \hbar\frac{\gc}{2}\sum_{\mu,\mu',\mu''}\left[\frac{1}{2}(1+d^2)\underbrace{(\opfwm^{\mu,\mu'\!,\mu''\!}_{\sym{,}\sym{,}\sym{,}\sym} + \opfwm^{\mu,\mu'\!\!,\mu''}_{\asym,\asym,\asym,\asym})}_{\mathrm{even\, and \, intra-band}}\right.\notag\\
& -d\sqrt{1-d^2}\underbrace{ (\opfwm^{\mu,\mu'\!,\mu''}_{\,\sym,\sym,\sym,\asym}+ \opfwm_{\sym,\asym,\sym,\sym}^{\mu,\mu'\!\!,\mu''\!} - \opfwm^{\mu,\mu'\!,\mu''}_{\,\sym,\asym,\asym,\asym} - \opfwm^{\mu,\mu'\!,\mu''}_{\,\asym,\asym,\sym,\asym})}_{\mathrm{odd\, and \, inter-band}}\notag\\
&\left. \notag + \frac{1}{2}(1-d^2)\underbrace{  (\opfwm^{\mu,\mu'\!,\mu''}_{\,\sym,\sym,\asym,\asym} + 4 \opfwm^{\mu,\mu'\!,\mu''}_{\,\sym,\asym,\sym,\asym} + \opfwm^{\mu,\mu'\!,\mu''}_{\,\asym,\asym,\sym,\sym})}_{\mathrm{even\, and \, inter-band}}\right].
\end{align}
\section{Numerical reconstruction of the dimer phase space}

\begin{figure*}
	\centering
	\includegraphics[width=0.9\linewidth]{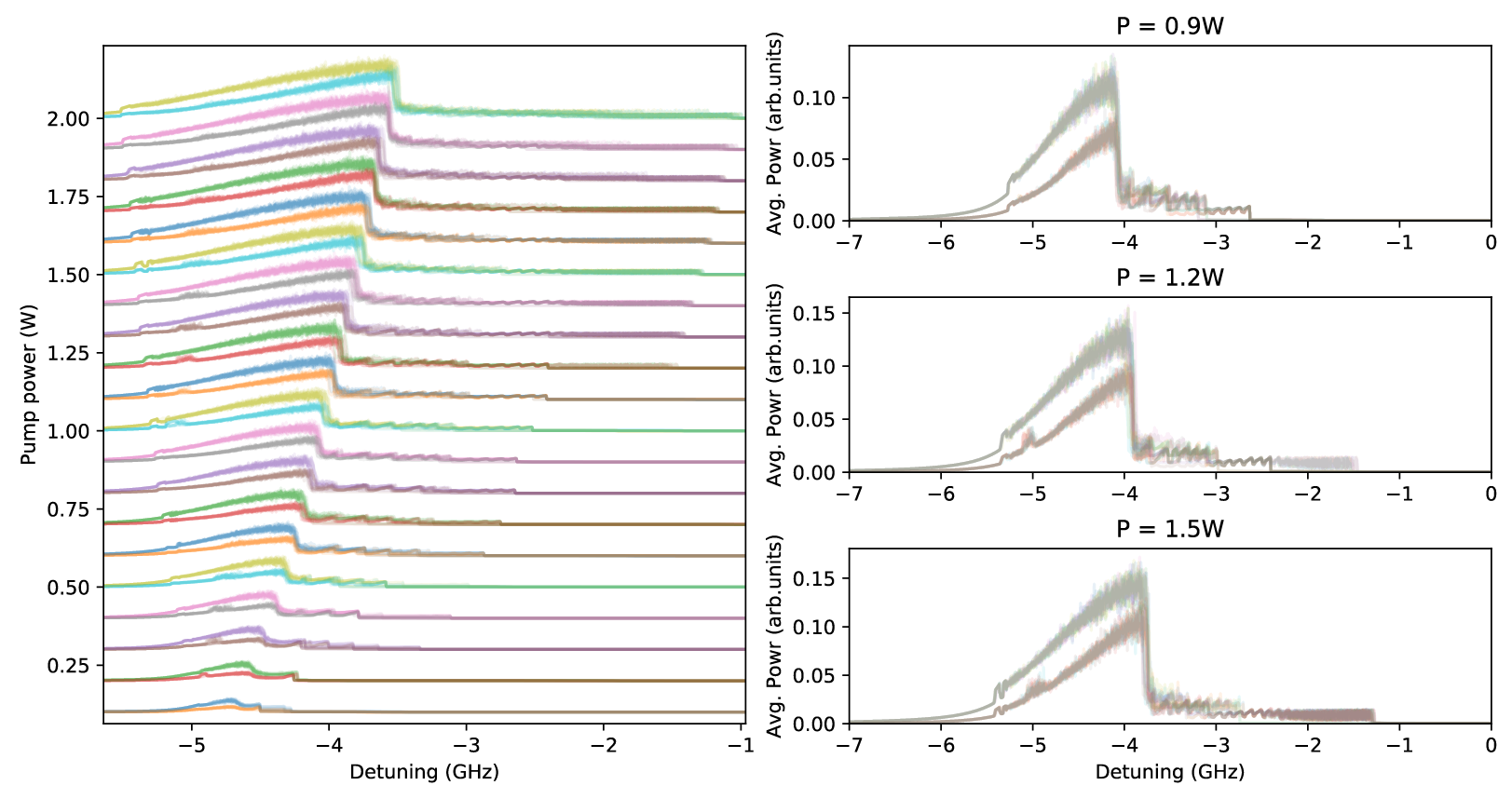}
	\caption{\textbf{Numerical reconstruction of the dimer phase space.} \textbf{(left)} Superposition of interactivity power traces. For every value of the pump power there 10 traces superimposed. \textbf{(right)} three examples for three different pump powers: 0.9, 1.2 and 1.5 W.}
	\label{fig:si1}
\end{figure*}

The phase diagram is reconstructed by superimposing 10 interactivity power traces and thereby identifying the range of parameter that corresponds to the diagram regions.
The coupling coefficient $J$ is considered linear and, at first, frequency independent. The angle between the geometrical center of resonators and the bus waveguide is chosen to be $\pi$ which corresponds to a simple vertical arrangement. The inter-resonator detuning $\delta$ is included by adding a correction to the integrated dispersion. The noise is taken on the level of $10^{-6}$ photons per mode with uniformly distributed random phases. Other simulation parameters are taken from the experimental measurements of real Si$_3$N$_4$ devices. Nonlinear propagation in the photonic dimer can be described by two coupled LLEs. Let $A(t,\theta)$ and $B(t,\theta)$ be slowly-varying complex field's envelops in the pumped and auxiliary resonators (shown in Fig.~1c (main article) by blue and red respectively). The moving frame is rotating with the same group velocity in both resonators but in opposite directions. We have chosen the following formulation of the coupled LLEs:
\begin{widetext}
\begin{equation}
\begin{split}
\frac{\partial A(t,\theta)}{\partial t}=&  -\left(\frac{\kappa_1}{2} +i\delta\omega\right)A + i\frac{D_2}{2}  \frac{\partial^2 A}{\partial \theta^2} + i g_\mathrm{K} |A|^2A +i J B(t,\theta') +\sqrt{\kappa_\mathrm{ex,1}} s_\mathrm{in} 
\\
\frac{\partial B(t, \theta)}{\partial t}=&  -\left(\frac{\kappa_2}{2} +i\delta\omega\right)B +i\frac{D_2}{2}  \frac{\partial^2 B}{\partial \theta^2} + i g_\mathrm{K} |B|^2B +i J A(t,-\theta'),
\label{eq:cLLe}
\end{split}
\end{equation}
\end{widetext}
where $\kappa_{1(2)} = \kappa_0+ \kappa_\mathrm{ex,1(2)}$ represent the total loss rate in the pumped and auxiliary cavities respectively composed of internal losses represented by $\kappa_0$ and losses due to the coupling $\kappa_\mathrm{ex,1(2)}$, $\delta \omega = \omega_0-\omega_p$ - pump laser detuning from the cavity resonance, $\theta \in [-\pi,\pi]$ - polar angle corresponding to the ring circumference, $D_2$ describes the deviation of the resonant frequencies from the equidistant grid defined by the FWM, $g_\mathrm{K}$ is the Kerr shift per photon,  $\theta' = \pi -2 \theta_C +\theta$ where  $\theta_C$ is an angle between centers of the rings and the bus waveguide. 
The phase diagram is numerically generated by exciting the system in a soft manner, i.e. adiabatically changing the laser detuning $\delta \omega$ from blue to red side of the AS hybridized resonance. The underlying simulations for the phase diagram are shown in Fig~\ref{fig:si1}.

\begin{figure*}
	\centering
	\includegraphics[width=0.99\linewidth]{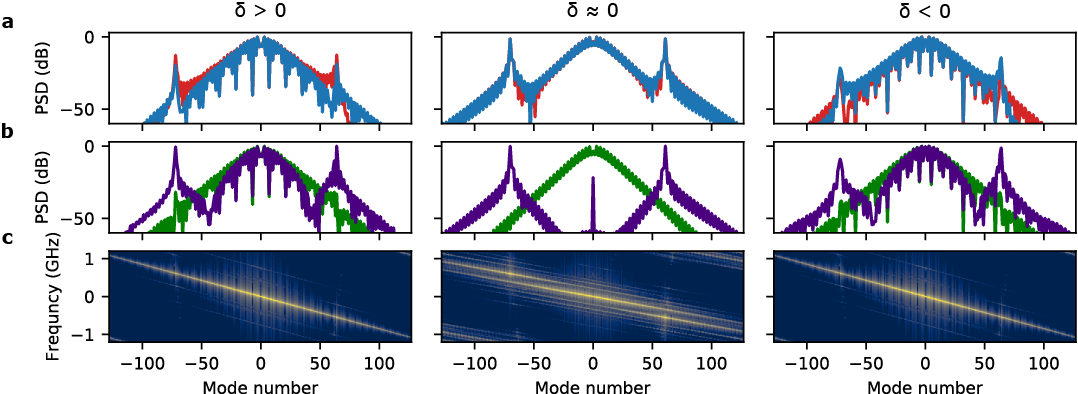}
	\caption{\textbf{Numerical investigation of the influence of the inter-resonator coupling dependence on the mode number.} Emergence of incommensurate dispersive waves is observed when the inter-resonator detuning is equal to zero which corresponds to the enhanced efficiency of the even inter-mode processes. \textbf{(a)} Power spectral density of the intraresonator field. \textbf{(b)} Supermode decomposition. \textbf{(c)} Reconstructed nonlinear dispersion relation.}
	\label{fig:si1_2}
\end{figure*}

The influence of the inter-resonator coupling dependence on the mode number is investigated numerically. The crucial role of this dependence can be seen in Fig.~\ref{fig:si1_2}. We simulated the photonic dimer dynamics for three coupling rates $\delta$. For the cases when $\delta$ is positive or negative and of the order of the coupling rate $J$, we observed commensurate dispersive waves appearing in the S supermodes. Change of the sign of  $\delta$ leads to the Fano shape inversion. When  $\delta$ is of the order of zero the incommensurate dispersive waves manifest themselves as parallel line to the main solitonic line in the reconstructed nonlinear dispersion relation Fig.~\ref{fig:si1_2}c.

\section{Photonic dimer fabrication and characterization}

Photonic dimers are fabricated with the photonic Damascene reflow process \cite{Pfeiffer2016Photonic}, deep-ultraviolet stepper lithography \cite{liu2018ultralow} and silica preform reflow \cite{pfeiffer2018ultra}. The ring radii are 125~$\mu$m, which results in a free-spectral range (FSR) of the microring resonators of $D_{1}/2\pi = 181.7$~GHz. Slight fabrication imperfections, related to lithography and the material removal rate occur over a much larger pattern size than the area of a photonic dimer and hence we observe excellent similarity of the individual microring resonators comprising the photonic dimer. Individual ring and bus waveguides are 1.5~$\mu$m wide and 800~nm high, which results in an anomalous second-order dispersion of $D_{2}/2\pi = 4.1$~MHz and third order dispersion of $D_{3}/2\pi = 4$~kHz. Inter-resonator coupling gaps vary from 300~nm to 450~nm. Conformal filling of narrow coupling gaps is facilitated by the photonic Damascene process \cite{Pfeiffer2016Photonic}. Both resonators are interfaced with bus waveguides to monitor the optical field in both resonators. Input and output coupling of light to and from the photonic chip is facilitated with double inverse tapers \cite{liu2018double} and lensed fibers.

\section{Linear characterization of photonic dimers}

\begin{figure*}
	\centering
	\includegraphics[width=0.75\linewidth]{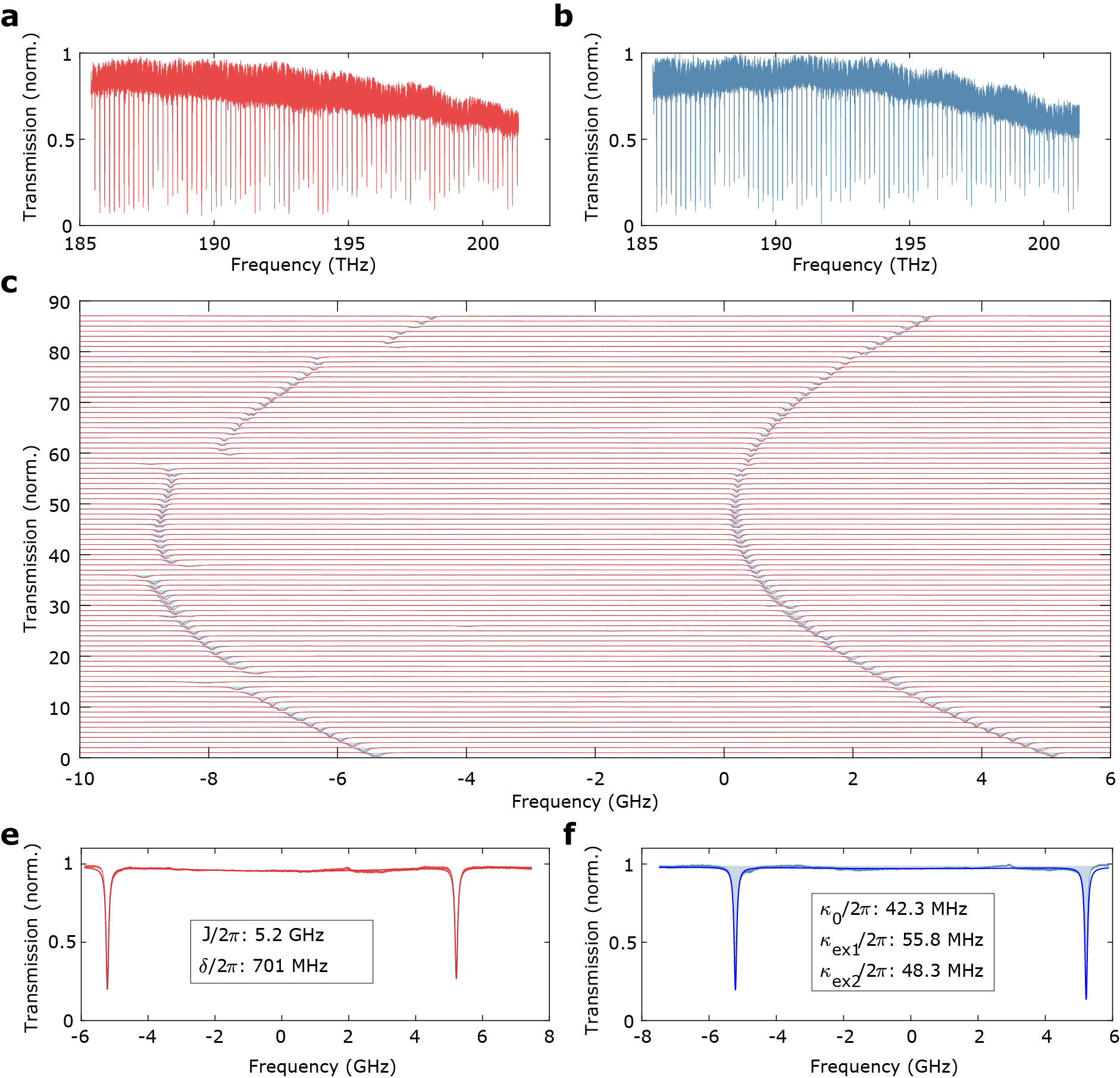}
	\caption{\textbf{Linear spectroscopy of photonic dimers}
	\textbf{(a,b)} Normalized frequency dependent transmission of top and bottom waveguides of a photonic dimer. \textbf{(c)} Superimposed Echellogram of photonic dimer normalized waveguide transmissions. Successive transmission lines are recessed by the cavity free spectral range of 181.8 GHz, starting at 186 THz in the bottom and ending at 203 THz in the top. Avoided mode crossings, i.e. scattering into transverse higher-order dimer modes, are much stronger on the symmetric dimer mode (S mode, left parabola) than on the anti-symmetric dimer modes (AS, right parabola). \textbf{(e,f)} Zoomed in transmission traces of top and bottom waveguide transmissions with fitted coupling, detuning and loss parameters.}
	\label{fig:si2}
\end{figure*}

We utilize the well established technique of frequency comb calibrated diode laser spectroscopy \cite{del2009frequency} to determine the linear properties of the photonic dimers, such as the cavity loss rates $\kappa$, the inter-resonator detuning $\delta$ and the linear coupling strength of the resonators $J$. Fig.~\ref{fig:si2}~a,b depicts the raw normalized transmission for both waveguides as a function of the optical frequency of the tunable diode laser. Residual reflection at the chip facets forms a Fabry-Perot cavity which induces a small sinusoidal modulation of the transmission traces. We use sequential fitting in a sliding window and subtract this modulation prior to further processing. In order to highlight the frequency-dependent properties of the photonic dimer resonances, we plot the transmission trace in a Echellogram (cf.~ Fig.~\ref{fig:si2}c), where the  frequency shift between subsequent vertical lines equals the cavity free-spectral range (FSR) of 180.75~GHz. Even without the application of external cavity tuners such as microheaters, we observe an interresonator detuning that is smaller than the evanescent coupling rate for at least 1/7 of the photonic dimers on the wafers, highlighting the excellent deposition and polishing uniformity of the photonic Damascene reflow process. It is well established that Si$_{3}$N$_{4}$ waveguides that feature anomalous dispersion also support multiple higher order modes, which are sufficiently confined in the waveguide core to interact with the fundamental mode and hence form so-called avoided mode crossings. We also regularly observe that coupling to higher order modes and the formation of avoided mode crossing is much stronger for the S-mode family than for the AS-mode family. A detailed study of this phenomenon and whether it is linked to a fundamental symmetry in the system is currently being conducted using finite element and finite difference time domain modeling and goes beyond the scope of the present study of soliton formation in the photonic dimer. 

\begin{figure*}
	\centering
	\includegraphics[width=0.75\linewidth]{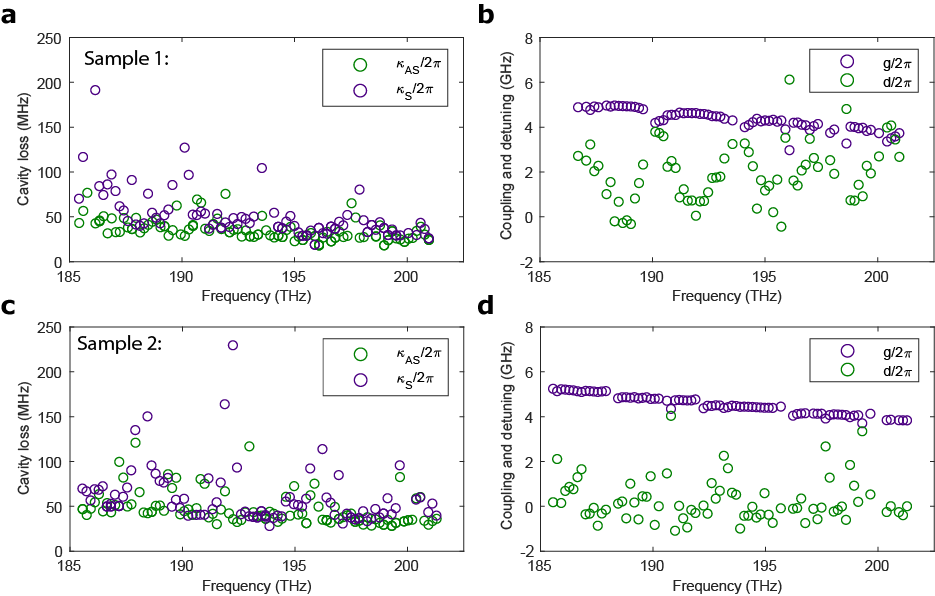}
	\caption{\textbf{Frequency dependent cavity dissipation and coupling rates of photonic dimers} Frequency-dependent results of photonic dimer fitting. a) Internal $\kappa_{0}/2\pi$ and external $\kappa_{ex}/2\pi$ loss rates for sample 1 and 2.}
	\label{fig:si3}
\end{figure*}

The relevant parameters are extracted by least-squares fitting of the model Eq.~\ref{eq:cLLe} in the adiabatic approximation ($\partial A / \partial t = 0$), neglecting the nonlinear term ($\gc = 0$). An exemplar fit of the resonance pair of the photonic dimer at 186.1~THz is plotted in Fig.~\ref{fig:si2}e,f. In general the evanescent inter-resonator coupling is frequency-dependent and periodic fluctuations of the fitted values of coupling $J$ and inter-resonator detuning are related to repeated modal crossings with the TE10 mode of the photonic dimer Fig.~\ref{fig:si3}.

\section{Experimental setup and data analysis}

\begin{figure*}
	\centering
	\includegraphics[width=0.75\linewidth]{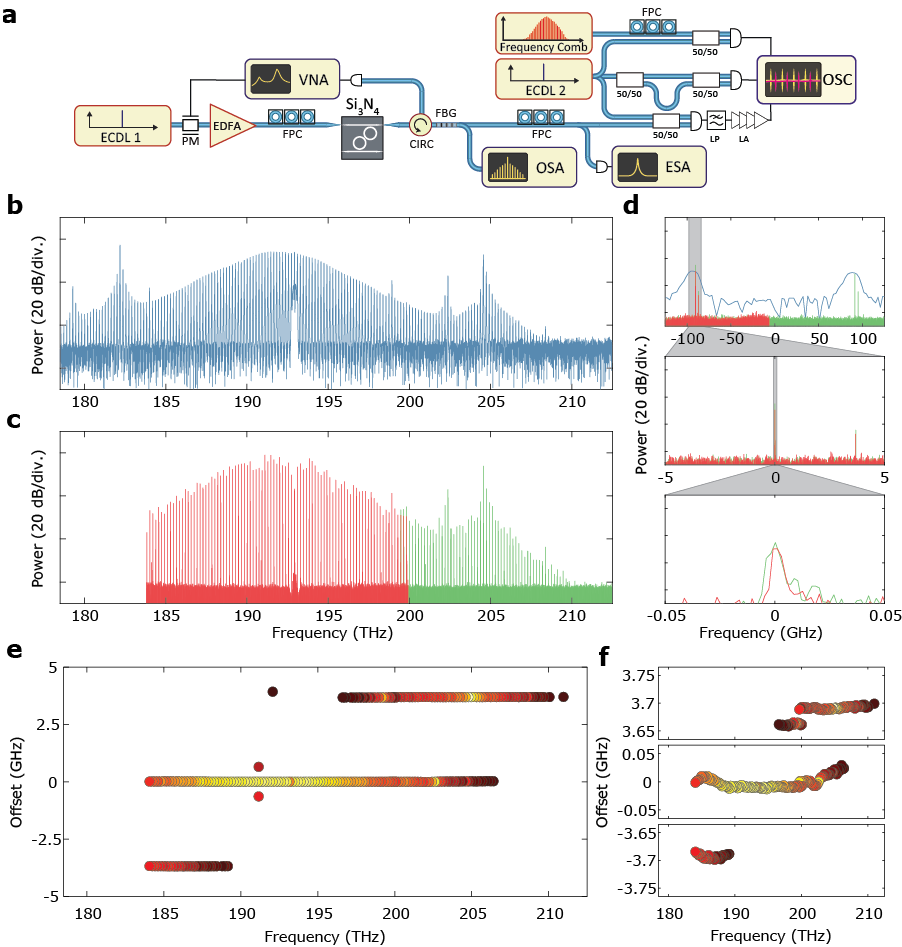}
	\caption{\textbf{Experimental setup for Kerr comb reconstruction of photonic dimer solitons}. a) External cavity diode laser (ECDL); phase modulator (PM); erbium doped fiber amplifier (EDFA); Fiber polarization controller (FPC); Optical circulator (CIRC); Vector network analyzer (VNA); Fiber Bragg grating (FBG); Optical spectrum analyzer (OSA); Electrical spectrum analyzer (ESA); low-pass filter (LP); Logarithmic amplifier (LA); Sampling oscilloscope (OSC) b) Optical spectrum measured with the grating-based optical spectrum analyser. c) Calibrated Kerr comb reconstruction trace of long wavelength(red) and short wavelength(green) ECDL scans. d) Zoom into the overlap region of the two laser scans. Grey area indicates spectral region highlighted in the panel below. e) Kerr comb reconstruction spectrogram. The offset frequency marks the frequency difference of the incommensurable dispersive wave from the soliton frequency comb. f) Zoom into offset regions around incommensurable dispersive waves (top and bottom) and soliton frequency comb. The spectral precision of Kerr comb reconstruction is limited by slow drifts of the pump laser and photonic dimer frequency.}
	\label{fig:si4}
\end{figure*}

\begin{figure*}
	\centering
	\includegraphics[width=0.75\linewidth]{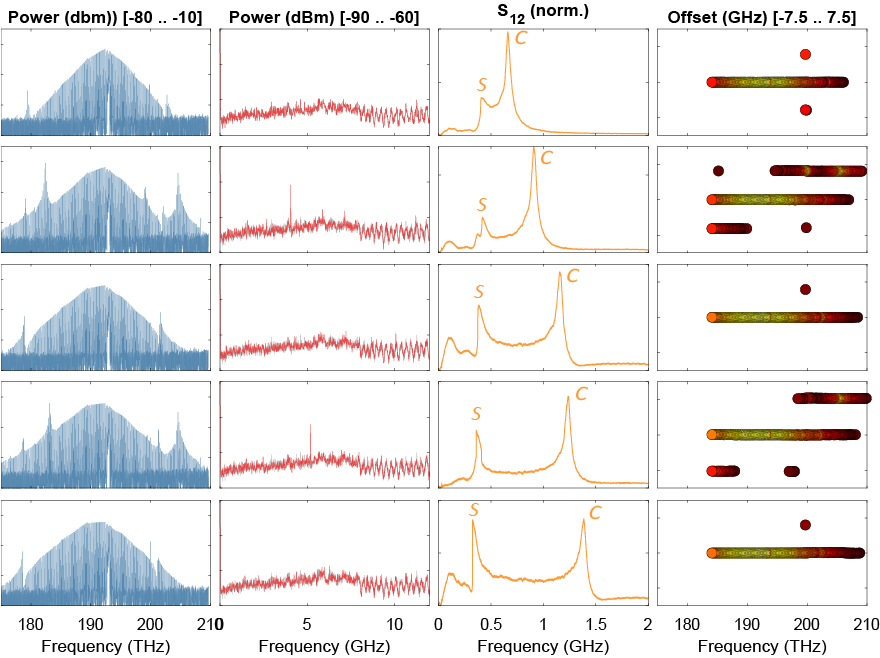}
	\caption{\textbf{Multi-soliton and photonic crystal states.}
		Optical spectra (blue), electrical spectrum of power fluctuations (red) with detector noise floor (grey) and phase modulation response (orange). $S,C$ resonance peaks result from the bistabilty of the soliton solution in the resonator. The detuning between the laser and the "hot" AS cavity mode is deduced from the position of the C-resonance peak. Physical units and y-Axis ranges are noted on top of the panels in the figure. The right panel depicts the results from Kerr comb reconstruction with a similar color scale as SI Fig.~\ref{fig:si4}.}
	\label{fig:si5}
\end{figure*}

The experimental setup for characterization of nonlinear frequency mixing and dissipative Kerr soliton generation in photonic dimers is depicted in Fig.~\ref{fig:si4}a. In order to avoid the emergence of incommensurate dispersive waves, we introduce unequal couplings to bus and drop waveguides of the dimer. It comprises a combination of the experimental setups used in Refs. \cite{herr2012universal} for Kerr comb reconstruction and \cite{guo2017universal} for dissipative soliton generation and phase modulation response measurements. A widely tuneable external cavity diode laser (ECDL) is passed through a fiber-coupled phase modulator (PM), amplified in an erbium doped fiber amplifier (EDFA) and coupled to the photonic chip. The laser is tuned into the antisymmetric resonance via the piezo tuning method described in Ref.~\cite{Herr2014Temporal}. Light from the chip is either retrieved at the transmission or drop waveguide ports. The pump light is reflected by a tuneable fiber Bragg grating (FBG), redirected by an optical circulator (CIRC) and impinges onto a fast photodiode (PD). The relative detuning of the pump laser and Kerr-shifted resonance is determined by interrogating the phase modulation response $S_{12}$ of the photonic dimer using the phase modulator (PM) driven by a vector network analyser (VNA) and receiving the signal from the PD. It is depicted in orange in Figs.~\ref{fig:si5},\ref{fig:si7}. The optical spectra generated by nonlinear interaction in the photonic dimer are analyzed in an optical spectrum analyzer (OSA) and depicted in blue in Figs.~\ref{fig:si5},\ref{fig:si7}. The intensity fluctuations of the light emerges either through chaotic modulation instability \cite{herr2012universal}, various breathing processes\cite{Lucas2017Breathing,Guo2017Intermode,yi2017single} or by mixing of components of soliton and incommensurable dispersive waves generated in the AS and S modes of similar mode number $\mu$, respectively.

In order to measure the nonlinear dispersion relation of soliton states and emergent nonlinear phenomena directly, we utilize the Kerr comb reconstruction method \cite{herr2012universal}, i.e. we reconfigure the frequency-comb assisted diode laser spectrograph \cite{del2009frequency,liu2016frequency} as heterodyne optical spectrum analyzer \cite{baney2002coherent} by superimposing the output of the photonic dimer with the scanning laser on a balanced photodetector. The spectral resolution is 4~MHz, determined by a 2~MHz low pass filter (LP). A high dynamic range and logarithmic response is achieved by inserting a multistage logarithmic amplifier (Analog Devices 8307) after the low-pass filter. Our data analysis work flow of the Kerr comb reconstruction is plotted in Fig.~\ref{fig:si4}b-d. In order to extend the spectral range of Kerr comb reconstruction, we employ two widely tuneable ECDL lasers with overlapping wavelength ranges (1500-1630~nm and 1350-1505~nm). Each laser scan takes about 20 seconds to complete and we continuously record the OSA and VNA signals in order to ensure that the Kerr comb reconstruction is performed consistently. The dissipative Kerr soliton itself can be used to precisely stitch the consecutive laser scans, which avoids the necessity of a stable transfer laser \cite{liu2016frequency}. The calibrated grating spectrograph (cf.~Fig.~\ref{fig:si4}b) is used to calibrate both the global frequency offset and the amplitude of the optical spectrum obtained by Kerr comb reconstruction (cf.~Fig.~\ref{fig:si4}c). As a result of heterodyne detection and the small resolution bandwidth, we obtain excellent dynamic range and noise floor (-90 dBm). In order to compress the vast spectral information, we perform a peak detection and record only the position and amplitude of distinct peaks in a scatter type plot (cf.~Fig~\ref{fig:si4}e,f). The precision of Kerr comb reconstruction is limited by drift of the ECDL laser pumping the photonic dimer. 

Figure \ref{fig:si5} depicts additional data complementing the results displayed in Fig. 3 of the main manuscript. The data is measured by coupling light in and out of the top waveguide, whereas the data in Fig. 3 of the main manuscript is obtained by coupling light in and out of the bottom waveguide. Successive measurements are recorded at gradually increased laser-resonance detuning in the AS resonator mode at 192.9~THz. 

\section{Multi-soliton and photonic crystal states}

\begin{figure*}
	\centering
	\includegraphics[width=0.75\linewidth]{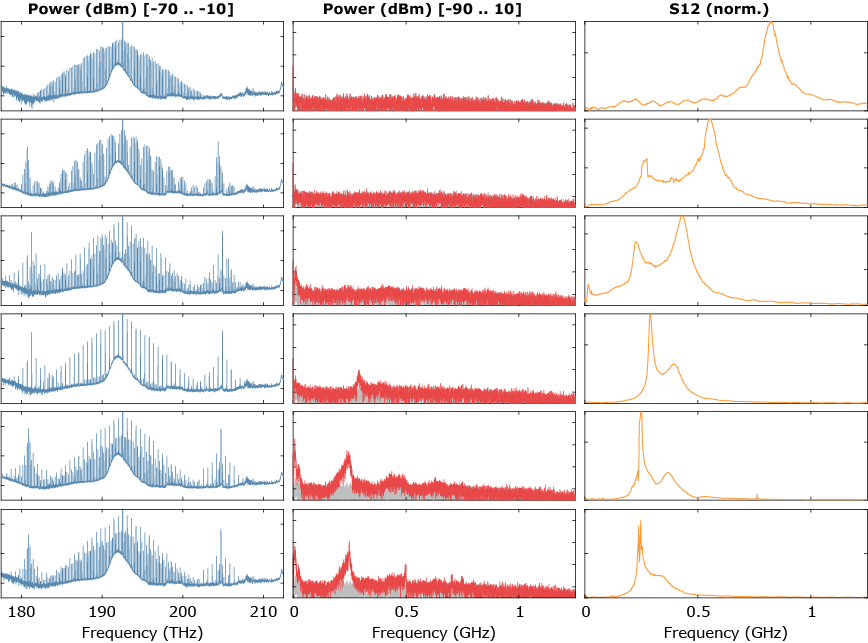}
	\caption{\textbf{Multi-soliton and photonic crystal states.}
		Optical power spectra (red, left), spectrum of intensity fluctuations (blue) with detector noise floor (grey) and phase modulation response (orange). Physical units and y-Axis ranges are noted on top of the panels in the figure.}
	\label{fig:si7}
\end{figure*}

The most surprising novel feature of the incommensurate dispersive waves is that they are directly linked to the solitonic state in the photonic dimer. This is most pertinently observed in multi-soliton and soliton crystal states. In Fig.~\ref{fig:si5} we depict the optical spectrum the amplitude noise spectrum and the phase modulation response of a series of multi-soliton and soliton crystal states, observed in sample 2 at moderate pumping power in the bus waveguide ($P_\mathrm{wg} = $360~mW). At low laser-cavity detuning $\zeta$, we observe both imperfect \cite{cole2017soliton} and perfect \cite{karpov2019dynamics} 4-soliton crystal states. Individual solitons can be eliminated by decreasing $\zeta$ and tuning the laser into the transient chaos regime of the soliton phase diagram \cite{karpov2019dynamics} or by increasing the detuning towards a resonant maximum of dispersive wave generation where the mutual interaction of the solitons via the strong dispersive waves in the S mode destroys one or more solitons. The top 4 panels of Fig.~\ref{fig:si7} show such a chain of multi-soliton states from a 4-soliton crystal to a single soliton state. All dispersive waves inherit the multisoliton spectral interference pattern, which is generally considered an argument for their coherence \cite{brasch2016photonic} and highlights that these incommensurate dispersive waves are generated by all solitons simultaneously. In case of the perfect soliton crystal we observe that the pattern is shifted by one mode on each side. In all cases, the spectral envelope of the incommensurate dispersive wave is a Lorentzian and not a Fano lineshape. This is due to the fact that the multi-GHz offset-frequency difference between soliton and dispersive wave leads to a temporal averaging of their constructive and destructive interferences on the detector.

\section{Breathing and intermode breathing soliton states}

Kerr comb reconstruction not only reveals the differences between the carrier-envelope-frequencies of the gear solitons and Lorentz-shaped incommensurable dispersive waves, it also proves the spectral coherence between gear solitons and the Fano-shaped dispersive waves and reveals spectral signatures of breathing and intermode breathing states directly. Both conventional \cite{Lucas2017Breathing} and intermode breathing \cite{Guo2017Intermode} are observed in the form of regularly spaced subcombs that span the full range of the soliton as well as the commensurable Fano-shaped dispersive waves. The appearance of intermode breathing occurs only on part of the investigated samples and is liked to the appearance of strong single mode dispersive waves that disturb the solitonic solution \cite{yi2017single}.

\begin{figure*}
	\centering
	\includegraphics[width=0.75\linewidth]{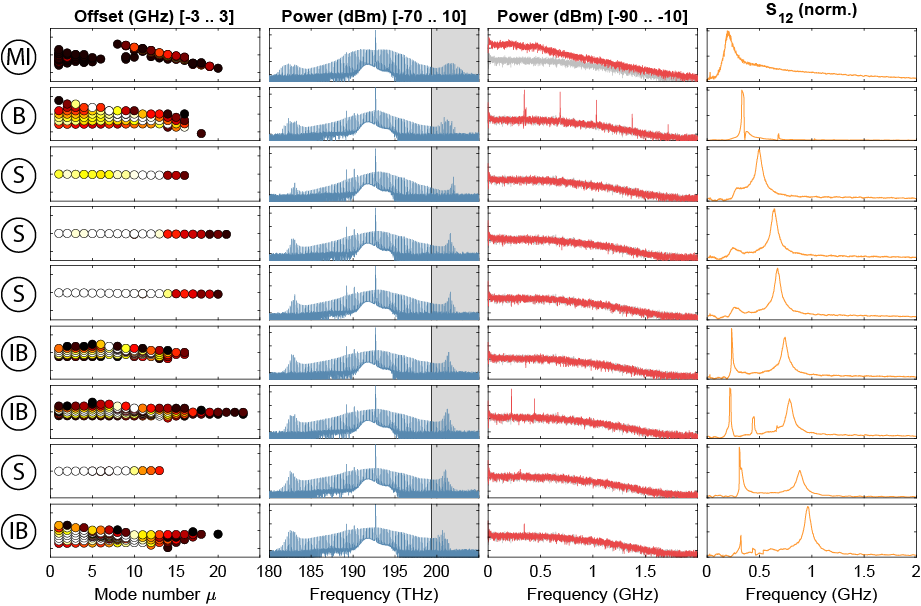}
	\caption{\textbf{Breathing and intermode breathing soliton states.} Optical spectra (red, left), electrical spectrum of power fluctuations (blue, middle) with detector noise floor (grey) and phase modulation response (orange). Physical units and y-Axis ranges are noted on top of the panels in the figure.}
	\label{fig:si8}
\end{figure*}

\bibliography{citations_SI}